\documentclass[prl,twocolumn,aps,amsmath,amssymb,showpacs]{revtex4}

\usepackage{graphicx}
\usepackage{dcolumn}
\usepackage{bm}
\usepackage{wrapfig}
\usepackage{amsmath,amsfonts,amssymb,bm,ulem}
\usepackage[usenames]{color}

\begin{document}


\title{Environmental Decoherence versus Intrinsic Decoherence}

\author{ P.C.E.
Stamp$^{\star}$ }

\affiliation{Pacific Institute of Theoretical Physics,
Physics \& Astronomy, \\
University of British Columbia, 6224 Agricultural Rd., Vancouver
B.C., Canada V6T 1Z1.
 }


\begin{abstract}

We review the difference between standard environmental decoherence
and 'intrinsic decoherence', which is taken to be an ineluctable
process of Nature. Environmental decoherence is typically modeled by
spin bath or oscillator modes - we review some of the unanswered
questions not captured by these models, and also the application of
them to experiments. Finally, a sketch is given of a new theoretical
approach to intrinsic decoherence, and this scheme is applied to the
discussion of gravitational decoherence.

\end{abstract}

\maketitle

\section{I. Introduction}

In this short paper I was given the job of briefly reviewing current
progress on the theory of environmental decoherence as applied to
condensed matter systems. In order to make the topic a little more
interesting I have added some material on 'intrinsic decoherence'.
This latter term denotes a process which is considered to be generic
to Nature, and which causes phase decoherence even without any
'averaging over an environment'. As such it is equivalent to a
breakdown of quantum mechanics. Many arguments have been given to
suggest that such a breakdown is something to be looked for --
perhaps the most compelling are those beginning either from the
clash between quantum mechanics and General Relativity, or from the
difficulties we have in thinking about macroscopic superpositions.

We will see that the topic of intrinsic decoherence is in its
infancy, and that it is then interesting to compare it to
environmental decoherence where things are a lot clearer. However in
what follows I will also try to emphasize some of the remaining
problems (as well as some of the successes), that we have with
environmental decoherence.

\section{II. Environmental Decoherence: Existing Models}

In order to say something useful about environmental decoherence,
let's begin by reviewing the current understanding of how this
works. This has come a very long way since the schematic models that
were used before the 1980's. One begins by writing the universe as a
Hamiltonian of form
\begin{equation}
H_{\mbox{\scriptsize eff}}^{osc} \;=\; H_o(P,Q) + H_{int}(P,Q;
p_{\nu}, q_{\nu}) + H_{env}(p_{\nu}, q_{\nu})
 \label{H-envD}
\end{equation}
where the $Q,P$ are generalized coordinates and momenta for the
central system of interest, and the $\{ q_{\nu}, p_{\nu} \}$ are the
same for the environmental degrees of freedom. The variety of models
for environmental decoherence then comes from:

\vspace{2mm}

(i) {\bf Models of the Environment}: At first glance it seems as
though we ought to require a large variety of different model
Hamiltonians for different environments. However with few
exceptions, it has been possible to reduce all of these to one of
two models, as follows:

\vspace{2mm}

{\it (a) Oscillator Baths}: Here we model the environment by a bath
of independent oscillators \cite{feynV63,cal83}, so that
\begin{equation}
H_{env}^{osc} \;=\;  \sum_{q = 1}^{N_o} ( {p_q^2 \over m_q} + m_q \omega_q^2 x_q^2 )  \\
 \label{H-osc}
\end{equation}
where the $\{ \omega_q \}$ are the frequencies of the $N_o$
oscillators. Opinion on oscillator bath models was polarized in the
early days - sometimes they were applied willy-nilly to almost any
environment. And yet in their original discussion of such models,
Feynman and Vernon \cite{feynV63} made it quite clear that the
oscillator bath representation was only generally valid if the
coupling between each environmental mode and the central system was
very weak. The most obvious case is when the environmental modes are
delocalized, extending over some large volume (so that $N_o$ is an
extensive quantity, depending in general on the UV cutoff $\Omega_o$
we employ). Examples are phonons, photons, magnons, electron-hole
pairs, and other collective excitations in a fermionic system. The
coupling is then taken to have the general form \cite{feynV63,ajl84}
\begin{equation}
H_{int}^{osc} = \sum_{q=1}^{N} [u_{q}(Q) x_{q} + v_{q}(P) p_q ]
 \label{hint-O}
\end{equation}
which is restricted to being linear in the oscillator coordinates
$x_q$ and their conjugate momenta $p_q$. The weakness of the
interactions $u_q(Q), v_q(P)$ is assured by the delocalization of
the modes (so that each mode has a normalization factor $\sim
N_o^{-1/2}$, and thus $|x_q|, |p_q| \sim N^{-1/2}$); in the
'thermodynamic limit' (ie., for large systems), when $N_o \gg 1$,
this works.

Actually this model is far more generally applicable than it seems.
Often couplings to oscillator modes that are higher-order in the
oscillator coordinates are important or even dominant in the real
world - but these can be lumped into linear couplings to effective
oscillators via temperature-dependent couplings. And while
anharmonic couplings between oscillators will obviously affect the
very long-time dynamics of the bath, they typically have little
effect on the system dynamics over experimentally relevant
timescales - which is what we are interested in. The basic idea
behind the oscillator bath model is one of linear response by the
bath modes, to the perturbation by the system - as such the idea
goes back at least to Rayleigh's model of friction \cite{rayleigh}.

\vspace{2mm}

{\it (b) Spin Baths}: Where oscillator bath models typically fail is
when the environmental modes are not delocalized. This is quite
common - examples include lattice defects, dislocation modes, spin
impurities, dangling bonds, nuclear spins, and localized
phonons/vibrons; there are many others. Because these modes are
localized in space, their Hilbert space is very small - most
commonly they can be modeled as a spin bath' of 2-level systems
\cite{PS00}, described by a set of $N_s$ `pseudospins' $\{
\mbox{\boldmath $\sigma$}_k \}$, with a Hamiltonian
\begin{equation}
H_{env}^{sp} \;\;=\;\; \sum_{k}^{N_s} {\bf h}_k \cdot
\mbox{\boldmath $\sigma$}_k \;+\; \sum_{k,k'}^{N_s} V_{kk'}^{\alpha
\beta} \sigma_{k}^{\alpha} \sigma_{k'}^{\beta}
 \label{H-SB}
\end{equation}
where each pseudospin feels a set of external fields $\{ {\bf h}_k
\}$, and inter-spin interactions $V_{kk'}$. These are then coupled
to the central system by an interaction:
\begin{equation}
H_{int}^{sp} \;=\; \sum_{k}^{N_s} \mbox{\boldmath $F$}_k(P,Q) \cdot
\mbox{\boldmath $\sigma$}_k
 \label{V-sp}
\end{equation}
Unlike the case of the oscillator bath, the interactions
$\mbox{\boldmath $F$}_k(P,Q)$ are not necessarily weak; there is no
reason to assume that $|\mbox{\boldmath $F$}_k(P,Q)| \sim
N_s^{-1/2}$ (indeed in many cases, $\mbox{\boldmath $F$}_k(P,Q)$ is
independent of $N_s$!), and moreover, the localized modes may also
be limited to some region in space, so that their number $N_s$ is
limited. We note also that one often must include the interactions
$V_{kk'}^{\alpha \beta}$ between the bath spins; even though they
are weak, they can have significant effects if $N_s$ is not large. A
key dimensionless parameter for a spin bath is $\lambda_k =
|\mbox{\boldmath $F$}_k(P,Q)|/|{\bf h}_k|$, which tells us the
relative strengths of the two fields acting on the bath spin
$\mbox{\boldmath $\sigma$}_k$. If $\lambda_k \gg 1$, the bath spin
dynamics is slaved to the central system; in many experimental
systems this is the case.

We see that the spin bath is inherently quantum-mechanical: discrete
environmental states in a finite Hilbert space have no classical
analogue. At low temperatures, such modes are everywhere in a solid
- even very pure and well-prepared metallic or semiconducting
crystal have a non-negligible concentration of defects and
dislocations, as well as paramagnetic impurities, and typically
their surfaces are very hard to control and will have surface
impurities, dangling bonds, and various kinds of lattice defect and
'charge fluctuator' modes, all of which can be described as spin
bath modes. And this of course is quite apart from nuclear spins,
which exist in profusion in almost all solids, and local vibrational
modes, which in many non-crystalline and organic systems play a key
role, coupling strongly to, eg., electronic modes, and behaving at
low $T$ very much like 2-level systems.

Reviews of both oscillator bath and spin bath models can be found in
the literature \cite{PS00,weiss99}.

\vspace{2mm}

(ii) {\bf Models for central systems}: These of course abound, and
theory has examined a large number of them, both for their intrinsic
interest, and in the context of experiments. The best known class of
models is one in which the central system is reduced to a 'qubit',
ie., a 2-level system with Hamiltonian
\begin{equation}
H_o \;=\; {\bf B}_o(t) \cdot \mbox{\boldmath $\tau$}
 \label{H-QB}
\end{equation}
where ${\bf B}_o(t)$ is some 'external' time-dependent field acting
on the qubit. Most theoretical analyses of decoherence for this
system have been done assuming a static ${\bf B}_o$. When coupled to
an oscillator bath, via a coupling $H^z_{int} = \sum_q u_q x_q
\hat{\tau}_z$, we have the diagonally-coupled 'Spin-boson' model
\cite{weiss99,ajl87}; it has been studied in the weak-coupling limit
ever since the beginning of quantum mechanics (notably in atomic
physics and quantum optics), and for more general couplings starting
in the late 1960's, in the context of RG theory and the Kondo
problem. One can also add a non-diagonal coupling $H_{int}^{\perp} =
\sum_q [u_q^{\perp} \hat{\tau}_+ H.c.] x_q$, which can be important.

When coupled to a spin bath, via a coupling $H_{int}^z = \sum_k {\bf
F}_k \cdot \mbox{\boldmath $\sigma$}_k \hat{\tau}_z$, the coupled
system is known as the 'Central Spin' model \cite{PS00}. Such models
were first studied in NMR and ESR, where again work was limited to
weak coupling; more general studies began with work on tunneling
magnetic systems \cite{magT}, and have accelerated with work on
decoherence in solid-state qubit systems \cite{PS00,sham}. Again,
one can generalize this to include non-diagonal terms, leading to a
general tensor interaction.

\begin{figure}
\vspace{-29mm}
\includegraphics[width=0.45\textwidth]{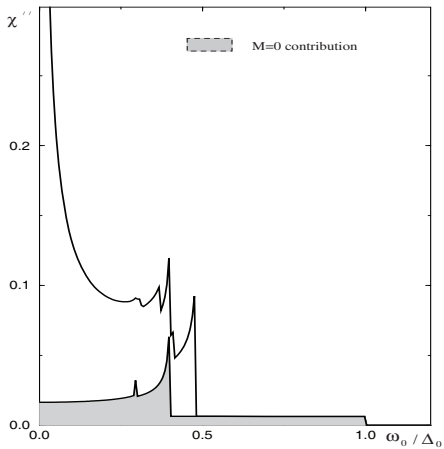}
\vspace{-9mm}
 \caption{Plot of $Im ~\chi_{zz}(\omega)$, the
absorption probability described in the text, for a qubit coupled to
a spin bath (the 'Central Spin' model). Without decoherence this
would be a sharp line at $\omega/\Delta_o = 1$. The shaded part
shows contributions from processes where the spin bath polarisation
is unchanged.}
 \label{chiF}
\end{figure}

The Spin-boson and Central Spin models have very different dynamics
\cite{PS00,ajl87}; this is inevitable, given the entirely different
structure of the Hilbert spaces for the two baths. The differences
are seen most easily in the lineshape (eg., in
$\chi''_{zz}(\omega)$, the Fourier transform to frequency space of
the time correlation $\langle \hat{\tau}_z(t) \hat{\tau}_z(0)
\rangle$). Suppose we hold ${\bf B}_o(t)$ constant in time. Without
the baths, $\chi''_{zz}(\omega)
 \propto \delta (\omega - \omega_o)$, where $\omega_o = 2B_o$.
As one increases the coupling to an oscillator bath, the
$\delta$-function for the spin-boson model first spreads into a
narrow Lorentzian, and then with increased coupling, the peak
broadens and moves to lower frequency; if the coupling becomes
strong enough (as in an Ohmic bath at strong coupling) the peak
moves to zero energy (critical damping), and eventually collapses to
a $\delta$-function at zero frequency (ie., localization in time).
On the other hand for a spin bath, the Central Spin lineshape
typically becomes very unconventional as one increases the coupling,
developing multiple peaks and looking anything but Lorentzian. The
corresponding behaviour in time also looks very different. Fig.
\ref{chiF} illustrates this for the Central spin model in a case
where the dimensionless coupling to the spin bath is intermediate in
strength - the contributions are divided between processes in which
the net bath polarization does not change, and those for which it
does. For more detailed discussion of such results see ref.
\cite{PS00}.

The reason for this difference is that (i) spins have a dynamics
very different from that of oscillators (the spin dynamics is highly
constrained compared to that of the oscillators), and (ii) in the
Spin-boson model the bath is always in the linear response regime -
each oscillator is only weakly affected by the qubit - whereas in
the Central spin model, the bath spin dynamics is strongly affected
by the qubit, and we are typically far from from a linear response
regime. If $\lambda_k \gg 1$, the bath spins simply follow the
central qubit (with deviations $\sim O(\lambda_k^{-1})$).

Another key difference between the Spin-Boson and the Central spin
model is related to this latter feature. In the spin-boson model
(and indeed in all oscillator bath models) dissipation and
decoherence are closely linked \cite{cal83}; one cannot have one
without the other, and both are connected to bath fluctuations via
the fluctuation-dissipation theorem. However in the Central spin
model it is not only possible but quite typical for strong
decoherence to occur with little or no dissipation. Indeed, the main
decoherence mechanism in the Central spin model is that of
'precessional decoherence', in which the bath spins precess in the
time-varying field of the qubit, without any energy exchange between
them \cite{PS00,SHPMP}. Again, this is a fairly generic feature of
spin bath decoherence.

For a more extended discussion of the differences between spin and
oscillator baths, see ref. \cite{SHPMP}.

Many other models for central systems have been discussed in the
literature, including tunneling systems, multiple qubit systems,
quantum critical systems, and quantum information processing
systems. We will only look at one, viz., a set of coupled qubits,
with Hamiltonian
\begin{equation}
H_o^{QIP} \;=\; \sum_j {\bf B}_j(t) \cdot \mbox{\boldmath $\tau$}_j
\;+\; \sum_{i<j} J_{ij}^{\alpha \beta}(t) \mbox{\boldmath
$\tau$}_i^{\alpha} \mbox{\boldmath $\tau$}_j^{\beta}
 \label{H-QIP}
\end{equation}
in which each local field ${\bf B}_j(t)$ on the $j$-th qubit is
subject to independent control, as are the couplings $J_{ij}^{\alpha
\beta}(t)$ between them. Then, to model decoherence, we couple each
individual qubit to the oscillator bath and the spin bath, using the
same local couplings as given above for a single qubit (but with, in
general, different couplings on each qubit). There are many possible
variants here - different qubits may couple to different spin baths
(being localized, these baths may be quasi-independent), and we by
changing the form of the coupling to oscillator baths (Ohmic,
super-Ohmic, etc.) we can radically alter the dynamics.

\begin{figure}
\includegraphics[width=6.5cm]{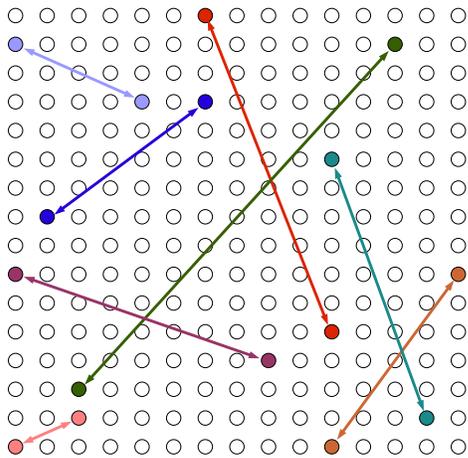}
\vspace{-2 mm} \caption{(Color on-line)  The dipolar interactions
between qubits can cause resonant flip-flop transitions over long
ranges - the figure depicts the qubits for which such a resonant
interaction might exist in some qubit network}.
 \label{dipole}
 \vspace{-3 mm}
\end{figure}

However a new kind of 'internal decoherence' enters with the
introduction of the inter-qubit interactions. Much depends on the
form of the $J_{ij}$ (in particular, whether they are long-range or
short-range in space); we will be interested particularly in
long-range dipolar interactions, which we can write in the form
$J_{ij} ^{\alpha \beta} \;=\; J_{ij}^0 \mathbb{D}_{ij}^{\alpha
\beta}$, where the interaction strength is $J_{ij}^0 \;=\; {\mu_0
\over \pi} \mu_B^2 / r_{ij}^3$ for spins separated by a distance
$r_{ij}$, and the spin tensor $\mathbb{D}_{ij}^{\alpha \beta}$ is
\begin{equation}
\mathbb{D}_{ij}^{\alpha \beta} \;=\; {g_i^{\gamma \alpha}
g_j^{\delta \beta} \over 4}
 \left[ \delta_{\gamma \delta} - {3 \over r_{ij}^2} r_{ij}^{\gamma} r_{ij}^{\delta}
 \right],
 \label{V_ijo}
\end{equation}
where the $g_i^{\gamma \alpha}$ are the effective "g-factor" tensors
for the qubits \cite{morello06}. Now these interactions are
'frustrating' \cite{parisi}, ie., they make it hard for the system
to order, and hard for simple pairwise interactions to operate
locally; and they are long-range. The radical effect that these
interactions can have is illustrated in Fig. \ref{dipole}, which
shows a set of typical 'resonating pairs' of qubits, able to perform
'flip-flop' transitions via the interaction $J_{ij}$. The key point
here is that in a 3-dimensional system, such pairs are just as
likely to be found far apart from each other as nearby (the decrease
in interactions strength being compensated by the larger number of
available pairs at longer range). Thus the quantum dynamics of the
system is not driven by short-range 'quantum diffusion' of
information - the information diffusion is spread all over the
system.

Why do we care about this? One reason is that it is very hard to get
rid of dipolar interactions like this - they are not easily
screened. So quantum computation is faced here with the problem that
unless one can actually use these interactions as part of the
computation protocol itself, then they cause 'correlated errors'
over long ranges. This is of course just another form of
decoherence. Thus we see that Hamiltonians like (\ref{H-QIP})
containing dipolar interactions describe a key aspect of all
qubit-based quantum information processing systems.

Study of problems like this on real systems has until recently been
limited to looking at the effect of both the dipolar interactions
and the coupling to the spin bath on, eg., the quantum phase
transition in these systems (see, eg., ref. \cite{LiHo}; note that
related problems have been discussed in the NMR literature
\cite{cory}). However in section IV we will discuss how it has now
become possible to quantify the various decoherence processes in
such a system.

\section{III. Environmental Decoherence: Some Questions}

A number of questions about the preceding framework jump out at us.
here are some:

(a) How good are these models? Thus - under what circumstances are
they valid, and how does one derive them for given systems? And when
can one use an oscillator bath representation even for localized
modes? What about models like 'noise' models, where we model the
environment by a noise source? And so on.

(b) Are there other kinds of decoherence (not captured by these
models)? What about decoherence caused by time variation of
parameters, or decoherence existing in the interaction channels -
neither of these is captured in the above. What about 'intrinsic
decoherence, ie., decoherence quite independent of any environment,
intrinsic to Nature?

(c) What about Experiment? Can we have large-scale quantum
computing, or large-scale quantum devices? What are the real limits
on quantum behaviour imposed by decoherence? And - is decoherence
inevitable at the macroscale?

(d) What about the 'Quantum Measurement Problem'? If there is such a
problem, does decoherence 'solve' it, or any of the other
foundational problems in quantum mechanics?

In the rest of this section we take a whirlwind ride through
questions (a) and (b). The next section will look at (c); and the
rest of the paper will take a closer look at intrinsic decoherence.
As for question (d), my general feeling is that until questions
(a)-(c) have been resolved, there is really nothing new that can be
said about (d).

\vspace{2mm}

{\it (a) How good are these Models?}: The derivation of the
oscillator bath models was discussed by Leggett et al.
\cite{cal83,ajl84}. The key here is the assumption of adiabaticity -
if we define instantaneous eigenstates $\tilde{\phi}_n (x_{\nu}, Q)$
and energies $\tilde{\epsilon}_n(Q)$ for the environment when the
central system has coordinate $Q$, then the oscillator energies
$\omega_q \equiv \omega_{nm} = \tilde{\epsilon}_n(Q) -
\tilde{\epsilon}_m(Q)$, and we can define the couplings $u_q, v_q$
in terms of the gauge potential
\begin{equation}
iA_{nm} = \prod_{\nu} \int d x_{\nu} \; \tilde{\phi}_n^*(x_{\nu})
{\partial \over \partial{Q}}  \tilde{\phi}_{m}(x_{\nu})
 \label{iA}
\end{equation}
provided that $Q(t)$ varies slowly enough in time that $\vert A_{nm}
\vert \ll \vert (\tilde{\epsilon}_n - \tilde{\epsilon}_m) \vert$. If
this condition is obeyed, then we can use the oscillator model even
if the original couplings to the bath are strong - the $\{ u_q,v_q
\}$ will still be small.

However in many cases there will be low-energy environmental modes
for which this condition is not obeyed (indeed some of them will
'resonate' with the central system, ie., their frequencies will
match the characteristic frequencies of dynamics of the central
system). In this case one must fall back on the perturbative
assumption discussed above, viz., that the the actual couplings to
the bath modes are weak. It is precisely this assumption that fails
in the case of many localized modes.

How good then are the spin bath models? In some cases the couplings
$\mbox{\boldmath $F$}_k(P,Q)$ and fields ${\bf h}_k$ are known
accurately from microscopic arguments and/or experiment (eg., with
hyperfine couplings to nuclear spins, in spin qubit systems). In
other cases we can sometimes only guess at these quantities (as with
the coupling to defects in magnets or superconductors). The key
assumption however is that the couplings $\mbox{\boldmath
$F$}_k(P,Q)$, or their rates of change, are small enough so that the
bath modes are not excited outside their restricted Hilbert spaces.
Note, incidentally, that we also assume that $V_{kk'}^{\alpha
\beta}$ is small compared to either $|{\bf h}_k|$ or
$|\mbox{\boldmath $F$}_k(P,Q)|$; otherwise the bath spins couple
together into delocalized modes, which behave more like oscillators.

It has quite common in some fields to approximate the effect of an
environment by coupling the system to an external 'noise' source,
ie., to approximate the full Hamiltonian in (\ref{H-envD}) by a form
$H(t) = H_o (P,Q) + V_N (Q;t)$, where the time-dependent potential
$V_N$ contains a random force acting on $Q$, and the effects of this
force are averaged over, using some technique (Langevin, Lindblad,
Fokker-Planck, etc.). But it well known that such models can never
fully capture the effect of integrating out a dynamical environment
\cite{feynH65}. They only describe the real part of the influence
functional, which itself completely characterizes the effect of the
dynamic environment on the system \cite{feynV63,feynH65}; moreover,
when approximated by, eg., Markovian processes, they also lose the
long-time correlations which are characteristic of the influence
functional of many environments. Physically, it is clear that a key
part of the environmental effect on the central system dynamics is
via a time-retarded 'back-reaction' on this system. We note that the
results may sometimes be quite counter-intuitive. The example of a
particle moving on a lattice while coupled to a spin bath
\cite{PS06} provides an example: the particle density matrix shows
simultaneous ballistic motion and anomalous diffusion, reminiscent
of weak localization. Further discussion of all this is quite
technical.

A related question is that of the crossover between spin bath and
oscillator bath models as one switches on the coupling between the
bath spins - when the interspin interaction dominates, we expect the
spin bath to behave like an oscillator bath. This crossover (which
for a macroscopic bath should under some circumstances become a
phase transition) has been discussed in some detail for the Central
Spin model \cite{PS00}, but there has been no general analysis as
yet.

\vspace{2mm}

{\it (b) What other kinds of Decoherence exist?}: A more obvious
question about failures of the oscillator and spin bath models is -
what do they leave out? This turns out to be an interesting line of
investigation. Here are a few suggestions (the list is not meant to
be exhaustive):

(i) Other environmental modes: It is certainly not clear that the
set of all environmental modes that might cause decoherence is
exhausted by oscillators and localized 'spin bath' modes. One
possible example, found in many systems at low energies, is that of
of delocalized quantum soliton modes - it is by no means clear (and
has certainly never been shown) that these can be described in all
cases by an oscillator bath model. There may also be other
possibilities

(ii) 3rd-party Decoherence: Since environmental decoherence arises
from the entanglement of a system with its environment, we can
certainly envisage situations where this happens without the direct
interaction of the two. For example, the state of each may be
'prepared' or otherwise conditioned by a 3rd party. A simple example
is that of the coupling between the vibrational and rotational modes
of a buckyball, on the one hand, and its centre of mass motion, on
the other, via the slits it passes through in a 2-slit experiment
\cite{SHPMP}. But one can easily imagine more subtle examples, and
the real question of interest here seems to be - how can we know
that a system has not previously entangled with some environment,
perhaps long in the past, perhaps via a 3rd party, in such a way as
to cause decoherence now? This seems to be to be an open question,
relevant in principle to many experiments.

(iii) Switching Decoherence: It is quite common, particularly in
quantum information theory, to write time-dependent Hamiltonians -
we have done so above in (\ref{H-QB}) and (\ref{H-QIP}). But this
raises two points. First, where is the time-dependent field coming
from that is causing this time variation, and shouldn't this source
term also  be included in the Hamiltonian (to take account of
possible back-reaction in the source)? And second - how do we know
that the time variation cannot project the system dynamics out of
the Hilbert space of the effective Hamiltonian we are using? This is
of course the whole problem in 'adiabatic quantum computation', and
in the similar problem of the dynamics of quantum phase transitions,
where a large number of energy gaps can decrease to very small
values during the time evolution. In this case, one can also include
all these extra levels in the 'environment' (now with time-dependent
energies). This topic is ripe for investigation, and certainly there
are many papers on the dynamics of driven systems coupled to a bath,
including some discussions of decoherence for simple models
\cite{carmichael}. However, to my knowledge no systematic study is
yet available.

(iv) Channel Decoherence: in quantum information theory it has been
common to discuss the decoherence arising during the transmission of
information down a channel. But such decoherence is ignored in most
of the effective Hamiltonians we write down, where coupling
constants are just treated as constants - this is equivalent to
assuming instantaneous interactions. The value of these interaction
constants depends on the UV cutoff $\hbar \Omega_o$ we assume in the
theory - but this takes no account, for a finite quantum computer,
of the time $T_o$ taken for signals to cross the computer. In
reality there is a set of timescales - the time $t_o$ taken for
qubits to switch, the time $\tau_o = 1/\Omega_o$ which is a UV
cutoff time, and the timescales associated with the qubit operations
themselves. Unless $T_o$ is very short compared to all of these, we
should not be assuming an instantaneous interaction, but rather a
retarded propagator. A proper treatment of this will be equivalent
to a microscopic derivation of channel decoherence.

(v) Intrinsic Decoherence: By intrinsic decoherence I will refer to
a decoherence in Nature that is {\it ineluctable}; it does not
result from averaging over any kind of external environment, and
thus can even exist for an isolated system. From this we see that it
must amount to a breakdown of quantum mechanics, because the linear
equation of motion on quantum mechanics preserves superpositions for
all time. The possible breakdown of quantum mechanics has been
investigated in many different ways in the last 40 years; some of
the best known efforts include the postulation of additional
non-linear terms \cite{weinberg89,nonL-S}, or else stochastic terms
\cite{GRW,Pearle}, to the Schrodinger equation, and a breakdown of
this equation caused by gravitational effects \cite{diosi,penrose};
this list is by no means exhaustive. In most cases the effect of
these deviations from standard quantum mechanics is to mimic
decoherence - in other words, in an experiment they would be
expected to add a further contribution to the existing environmental
decoherence, which may be very hard to distinguish from it.

It is fair to say that our understanding of all five of the above
decoherence mechanisms is more or less open right now - and all of
them are interesting and potentially important for experiments. In
the second half of this paper I will be looking at just one of them,
viz., intrinsic decoherence. However, a key challenge for any
experiment looking for intrinsic decoherence will be to distinguish
all sources of environmental decoherence from any intrinsic
decoherence (and if possible eliminate these sources). Therefore we
first take a look at how it is possible to nail down environmental
decoherence sources in experiments (cf. question (c) above).

\section{IV. Environmental Decoherence: Theory vs. Experiment}

In the last decade many experiments have been done looking at
decoherence in both quantum optics and condensed matter systems - in
the last couple of years these have been extended to hybrid systems
like optomechanical, optomagnetic, or opto-superconducting systems,
in which some combination of optical cavities, superconducting or
magnetic qubits, and mechanical or electromechanical oscillators are
coupled and put collectively into a quantum state. In some of these
experiments the decoherence mechanisms were fairly simple, involving
coupling of an oscillator degree of freedom to photons or phonons,
and in this case the experimental decoherence rates did not wildly
disagree with simple theory. In more complicated cases, usually
involving larger systems where the environment involved spin bath
degrees of freedom, the experimental decoherence rates have often
been much larger than naive theory would predict.

The temptation is to simply write these discrepancies off as 'junk'
effects, coming from some combination of defects, impurities, etc.
However the problem here is that if the experiments are supposed to
be testing, eg., the applicability of quantum mechanics to large
and/or complex systems, such a temptation is no longer open to us -
the theory should then also be predicting the environmental
decoherence rates, and a genuine experimental test of quantum
mechanics will then be testing such predictions.

It is of course always possible to argue that the theoretical models
are perfectly OK, but that we are analyzing their dynamics
incorrectly. All theoretical predictions assume that we can average
over all environmental modes to get a reduced density matrix for
central system alone; the dynamics of this reduced density matrix is
then described by a propagator $K(2,1) = K(Q_2,Q_2^{\prime}; Q_1,
Q_1^{\prime}; t_2, t_1)$, so that the reduced density matrix
$\bar{\rho}(2) = \bar{\rho}(Q_2,Q_2^{\prime};t_2)$ evolves according
to
\begin{equation}
\bar{\rho}(2) \;=\; \int d1 K(2,1) \; \bar{\rho}(1)
 \label{K-rho}
\end{equation}
where $d1$ means $\int dQ_1 dQ_1^{\prime}$. The propagator can be
written in the general path integral form \cite{feynV63}
\begin{equation}
K(2,1) \ = \displaystyle \int^{Q_2}_{Q_1} {\cal D}Q \displaystyle
\int^{Q^{\prime}_{2}}_{Q^{\prime}_{1}} {\cal D} Q^{\prime} \ e^{{-i
\over \hbar} (S_o[Q]  -  S_o[Q^{\prime}])} {\cal F}[Q,Q']
 \label{3.3}
\end{equation}
where $S_o[Q]$ is the free central system action, and ${\cal F}[Q,
Q^{\prime}]$ is the ``influence functional", defined in general by
\begin{equation}
{\cal F}[Q,Q'] =  \prod_{\nu} \langle \hat{U}_{\nu}(Q,t)
\hat{U}_{\nu}^{\dag}(Q',t) \rangle \;, \label{neto.2}
\end{equation}
Here the unitary operator $\hat{U}_{\nu}(Q,t)$ describes the
evolution of the $\nu$-th environmental mode, given that the central
system follows the path $Q(t)$ on its "outward" voyage, and $Q'(t)$
on its "return" voyage; and ${\cal F}[Q,Q']$ acts as a weighting
function, over different possible paths $(Q(t),Q'(t'))$.

The calculation of the influence functional is not necessarily easy.
It is given in general form by a functional integral over the
environmental modes, as
\begin{equation}
{\cal F}[Q,Q'] = \prod_{\nu} \int {\cal D} q_{\nu} \int {\cal D}
 q_{\nu}^{\prime} \exp {i \over \hbar} \Psi(q_{\nu},
 q_{\nu}^{\prime}; Q,Q')
 \label{F-osc}
\end{equation}
where the complex phase $\Psi(q_{\nu}, q_{\nu}^{\prime}; Q,Q')$ is
given by the difference in actions over the two paths in the reduced
density matrix, ie.,
\begin{equation}
\Psi(q_{\nu},
 q_{\nu}^{\prime}; Q,Q') = S_{B}(q_{\nu},Q) - S_{B}(q_{\nu}^{\prime}, Q')
 \label{Psi}
\end{equation}
The 'bath action' $S_{B}(q_{\nu},Q)$ just sums over the
environmental and interaction terms, ie., we have $S_{B} = \int
d\tau L_B$, where
\begin{equation}
L_{B}(q_{\nu},Q) =  L_{int}(q_{\nu},Q) + L_{env}((q_{\nu})
\end{equation}

Everything then depends on the form taken by the influence
functional. The oscillator bath influence functional was derived by
Feynman and co-authors \cite{feynV63,weiss99}; one lets $q_{\nu}
\rightarrow x_q$, and the Lagrangian $L_B (x_q,Q)$ is then
\begin{equation}
L_q (x_q, \dot{x}_q; t) \;=\; {m_q \over 2}[\dot{x}_q^2 - \omega_q^2
x_q^2] - \gamma_q (t) x_q
 \label{L_q}
\end{equation}
where $\gamma_q(t)$ is the driving force acting on the $q$-th bath
oscillator, coming from the central system. Thus, eg., if the
coupling in (\ref{hint-O}) is $L_{int} = -\sum_q F_q(Q) x_q$, we
have $\gamma_q(t) = F_q(Q(t))$. The bath action $S_B$ is then a
quadratic form in the oscillator coordinates and one can find an
exact expression for ${\cal F}[Q,Q']$. However, depending on the
form of $F_q(q)$ as a function of $Q$, the final integrals in
(\ref{3.3}) may still be highly non-trivial. This topic has been
studies at great length - a good review is by Weiss \cite{weiss99}.

The spin bath influence functional is very different, and has been
studied much less. Now one lets $q_{\nu} \rightarrow \mbox{\boldmath
$\sigma$}_k$, and the Lagrangian $L_B (\mbox{\boldmath
$\sigma$}_k,Q)$ takes the form
\begin{equation}
L(\mbox{\boldmath $\sigma$}_k, \dot{\mbox{\boldmath $\sigma$}}_k; t)
 \;=\; {\boldmath{\cal A}}_k \cdot {d \mbox{\boldmath $\sigma$}_k \over d\tau} -
 \mbox{\boldmath $\Upsilon$}_k(t) \cdot \mbox{\boldmath $\sigma$}_k
 \label{L_k}
\end{equation}
where ${\boldmath{\cal A}}_k$ is formally the static gauge field
which is produced by a monopole of unit strength at the centre of
the Bloch sphere for $\mbox{\boldmath $\sigma$}_k$ (and which leads
to the Berry phase for $\mbox{\boldmath $\sigma$}_k$
\cite{berry84}). The time-dependent field $\mbox{\boldmath
$\Upsilon$}_k(t)$ is given by
\begin{equation}
 \mbox{\boldmath $\Upsilon$}_k(t) \;=\; {\bf h}_k + {\bf F}_k(t) + \mbox{\boldmath $\xi$}_k (t)
 \label{b_k}
\end{equation}
in which ${\bf h}_k$ is the static external field acting on the
$k$-th bath spin, and ${\bf F}_k(t)$ is the dynamic field acting on
this spin, coming from from the central system ({\it cf.} eqtns.
(\ref{H-SB}) and (\ref{V-sp})). The extra $t$-dependent fluctuating
field $\mbox{\boldmath $\xi$}_k (t)$ is defined by its components as
\begin{equation}
\xi_k^{\alpha} (t) = \sum_{k'} V_{kk'}^{\alpha \beta}
\sigma_{k'}^{\beta}(t)
 \label{xi_k}
\end{equation}
and to first approximation this can be dropped, since
$V_{kk'}^{\alpha \beta}$ is a small perturbation on the field ${\bf
h}_k + {\bf F}_k(t)$.

This Lagrangian is much harder to deal with than the oscillator
Lagrangian - indeed the solution to the problem of a spin-$1/2$ in
almost any but the simplest time-dependent fields is practically
impossible to determine (something which always surprises students).
In spite of this, detailed expressions have been produced for the
dynamics of systems coupled to spin baths, which are believed to be
accurate in a fairly wide parameter range, at least up to a certain
time \cite{PS00,sham}. Nevertheless, given the uncertainties here,
one obvious way that one might try to explain the discrepancies
between theoretical and experimental environmental decoherence rates
in so many experiments, is simply to argue that the theoretical
calculations of spin bath decoherence rates are just way off target.

What is obviously required here is a confrontation between theory
and experiment where one has accurate knowledge of the physical
sources of the decoherence (notably the spin bath modes). In what
follows I would like to briefly describe one specific experiment,
which did explicitly attempt such a test. The experiment looked at a
macroscopic crystal of tunneling magnetic molecules, and
specifically was looking for the macroscopic spin polarization wave
that should propagate in a region of the sample if the decoherence
rate is low. The experiment is interesting because it checks out not
just decoherence from local coupling to both oscillator and spin
baths, but also a more insidious decoherence which comes from the
long-range dipolar coupling between all of the tunneling molecular
spins.

\begin{figure}
\includegraphics[width=0.35\textwidth]{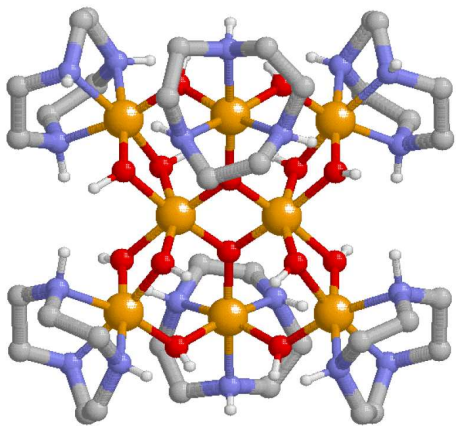}
\caption{Simplified ball-and-stick representation of the
[Fe8O2(OH)12(tacn)6]8+ system (the "Fe-8" molecule), showing the
octahedral iron ions, the oxo and hydroxo bridges and the
1,4,7-triazacyclononane (tacn) ligands completing the coordination
spheres.} \label{Fe8}
\end{figure}

The experiment was done on a crystal of "Fe-8" molecules; each
molecule contains 8 Fe ions in a small core, which are surrounded by
a complicated sheath of ligand molecules, containing H, C, N, O, and
Br ions (see Fig. \ref{Fe8}). The spins in the central magnetic core
are locked together by superexchange interactions into a spin-10
system, whose dynamics is governed by a biaxial crystal field.
Depending on the different isotopic concentrations in the molecules,
each one contains up to 216 nuclear spins. For the discussion here
we can just imagine that at sufficiently low temperature (below
about $2.5~K$), we have a set of spin qubits, each of which couples
locally to both a phonon oscillator bath and a nuclear spin bath.
These couplings, including all of the hyperfine couplings to the
different nuclear spins, are very well understood quantitatively.
The same is true of the dipolar coupling between the qubits, and so
we can model the system very accurately by the effective Hamiltonian
$H = H_o + H_{SB} + H_{osc}$, where
\begin{equation}
H_o \;=\; \sum_j \Delta_o({\bf H}_{\perp}) \hat{\tau}_x + \sum_{i<j}
J_{ij}^{\alpha \beta}({\bf H}_{\perp}) \hat{\tau}_i^{\alpha}
\hat{\tau}_j^{\beta}
 \label{HoFe8}
\end{equation}
sums over all crystal lattice sites, and where $H_{SB}$ and
$H_{osc}$ describe the nuclear spin and phonon terms respectively.
The temperature $T$ of the system is defined by the phonons; there
is also an magnetic field ${\bf H}_{\perp}$ applied in such a
direction that we can control the 'tunneling splitting' $\Delta_o$
over a very wide range (some 7 orders of magnitude), without adding
any terms $\propto \hat{\tau}_j^z$. The effective interaction energy
$J_{ij}^{\alpha \beta}({\bf H}_{\perp}) \hat{\tau}_i^{\alpha}
\hat{\tau}_j^{\beta}$ is actually dominated by longitudinal terms
$J_{ij}^{zz}$, so $H_o$ then becomes the famous dipolar quantum
Ising model. If the interaction were short-ranged, with strength
$J_o$, we would get a standard zero-$T$ quantum phase transition
when $|\Delta_o/J_o| \sim O(1)$. However the interaction here is
dipolar, and moreover we also have the couplings to the oscillator
and spin baths - this changes things fundamentally.

This system is a key test case for quantum computing, for the
following reason. We expect almost any solid-state gate-based
quantum computer to possess local couplings between the qubits and
spin and oscillator baths; and moreover, we expect long-range
dipolar interactions to exist between the qubits, either electric or
magnetic (or even strain-mediated). Such long-range interactions are
very dangerous for quantum computation, because conventional error
correction routines do not work for them - indeed it is not yet
clear how one would correct errors arising from them, although there
are arguments that concatenated error correction could do the job
\cite{kitaev06}. In any case, we see that this model contains all of
the agents of decoherence that we may expect to arise in a
multi-qubit quantum computer or quantum register (apart perhaps from
the switching decoherence and channel decoherence noted above). In
this sense the experiment to be described is the first one to test
all of these decoherence mechanisms together. A key feature of this
experiment was that all the hyperfine interactions to the nuclear
spin bath were known accurately, and so one can in principle
directly test the theory of spin bath decoherence, and the
calculational methods used to derive decoherence rates.

\begin{figure}
\vspace{-16mm}
\includegraphics[width=8.5cm]{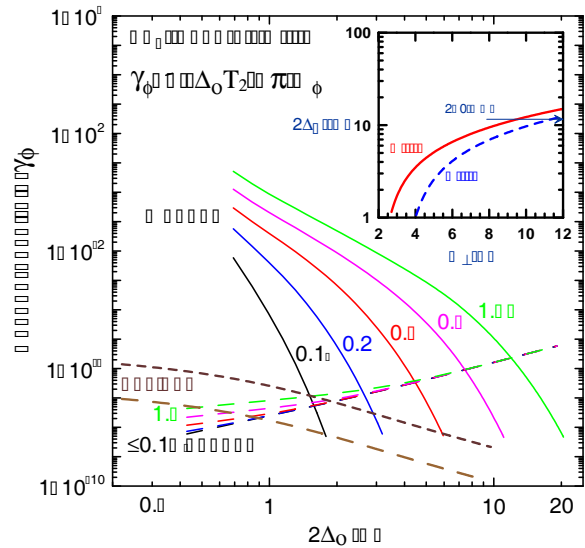}
\vspace{-15mm} \caption{(Color on-line)  Dimensionless spin
decoherence rates as a function of the qubit splitting $2\Delta_o$
for a crystal of Fe-8 molecules, with a field ${\bf H}_{\perp} ||
\hat{x}$ along the $x$-axis. Brown dashed line - nuclear
contribution to decoherence rate (upper line - protons at H sites;
lower line - deuterons at H sites). Solid lines - magnon
contributions at different temperatures, caused by intermolecular
dipolar interactions. Long dashed lines - phonon contributions
corresponding to the same temperatures as in the magnon case. Inset:
the splitting $2 \Delta_o$ as a function of applied transverse
field}.
 \label{decoR}
 \vspace{-10mm}
\end{figure}

The theory for the decoherence rate as a function of temperature,
applied field, and for different isotopic concentrations of
different nuclear spins, was actually carried out in 2006, and gave
detailed predictions for the Fe-8 system \cite{morello06}, in a spin
echo configuration suggested in that paper. The basic idea was that
one could measure the '1-spin' decoherence rate by doing a global
spin echo experiment, in which the entire spin system was set into a
spatially uniform precessional motion. One would of course like to
measure the dynamics of multi-spin density matrices, and this will
be a key goal for the future. In any case, one finds theoretically
that the dimensionless decoherence rate (defined as $\gamma_{\phi} =
1/T_2 \Delta_o$, where $T_2$ is the 1-spin decoherence time
extracted from spin echo measurements) should vary enormously
depending on the field (which controls $\Delta_o$), and on the
temperature (see Fig. \ref{decoR}). The phonon decoherence rates
increase quickly as the qubit tunneling frequency $\Delta_o$
increases (because the phase space for phonon emission is
increasing). On the other hand the nuclear spin bath decoherence
rates were predicted to fall rapidly at high $\Delta_o$ because the
nuclear spins can no longer follow the qubit dynamics. Thus for a
single qubit, one arrives at an optimal $\Delta_o$ (the 'coherence
window' \cite{tup04}), where the decoherence will be at a minimum.
However the dipolar decoherence, mediated by magnons, messes this up
completely - and the only way to suppress it is by going to low
temperatures (the qubit flip-flop processes which excite the
magnons, and which are caused by dipolar interactions, are gapped in
energy; they are then exponentially suppressed at low $T$).

A number of attempts were made by different experimental groups to
test this theory, but they were rendered very difficult by the
strong dipolar decoherence, which forced a very low-$T$ experiment
when conventional fields were used. This problem was circumvented by
Takahashi et al. \cite{taka11}, who went to fields in excess of
$11~T$, and also changed the field configuration. A very
satisfactory agreement with theory was then found, with no
adjustable parameters in either the theory or the experiment. This
experiment actually accomplished some other notable goals - it was
the first experiment to see macroscopic spin precession in an
ordered array of qubits, and also achieved a lower decoherence rate
than any previous molecular spin qubit system (the decoherence
"Q-factor" or inverse decoherence rate, went to $1.5 \times 10^{6}$,
and if these experiments had gone to lower $T$, it should have gone
to $6 \times 10^7$). However what interests us here is that it
tested all three decoherence mechanisms simultaneously (including
the first ever measurement of dipolar decoherence). This is because
by varying $T$ and ${\bf H}_{\perp}$, one can independently vary the
three mechanisms). So the obvious question is - what was being
tested in this experiment?

The two most obvious conclusion from the results are (i) that since
apparently all three decoherence mechanisms checked out, there could
have been only negligible 'extrinsic' effects in the experiment - no
hidden decoherence from dirt effects like hidden defects, dangling
bonds, etc., so that the only spin bath modes were the nuclear
spins; and (ii) that the calculations of both spin bath decoherence
and dipolar decoherence are actually reliable. To this we can add
that the agreement with experiment for phonon decoherence was to be
expected - this is a well-understood process. The success of the
predictions for spin bath and dipolar decoherence is welcome - for a
long time it was not obvious if disagreements between theory and
experiment came from experimental dirt or inadequate theory.

But let us now return to the question (c) posed in section 3. What
can we now say about decoherence for large-scale quantum processes,
including the kind that would be involved in quantum computing with
many qubits? Indeed, to what extent can one say that in doing
experiments like this we are testing fundamental theories of quantum
mechanics on large systems, in the same way that experiments on,
say, coherence in superconducting SQUIDs \cite{chiorescu03} are
supposed to be doing?

It seems to me that there are two basic approaches that one can take
to this question. One is to probe the predictions of quantum
mechanics in situations where its results are very
counter-intuitive, and apparently violate deep prejudices we have
about Nature. Examples of this approach are the inequalities derived
by Bell \cite{bell64}, which probe non-local entanglement, and have
now been tested in many different ways on microscopic systems (and
indeed constitute a 'resource' for quantum computation); and the
generalization of these to time inequalities, to discuss possible
experiments testing 'macroscopic realism' on superconducting SQUIDs
\cite{gargL85}. It is clear that the experiments described above do
not yet test quantum mechanics at the macroscopic level in this sort
of way, but they certainly could be adapted to do so, and probably
will be in the near future.

However a second way of testing quantum mechanics, rather than just
probing its consequences, is to develop alternative theories, and
then do experimental tests which compare these directly with quantum
mechanics, and indeed distinguish explicitly between them. This is
precisely what theories of intrinsic decoherence aim to do, and so
we now turn to these.

\section{V. Intrinsic Decoherence: Theoretical Framework}

As noted earlier, by 'intrinsic decoherence' we mean a process which
looks like decoherence (ie., one in which phase coherence or phase
interference is destroyed with time) but where this process is
intrinsic to Nature, ie., not obtained by simply averaging over
'external' degrees of freedom that happen to be entangled with the
system of interest. As such, an investigation into intrinsic
decoherence amounts to a search for a breakdown of quantum
mechanics, of a kind that mimics (at least up to a point)
conventional environmental decoherence. Thus we are looking for a
breakdown of quantum mechanics, of a specific kind.

For many physicists the first reaction to this idea is probably to
ask why anyone would bother looking at all! The two main reasons are
(i) the well-known difficulties attending macroscopic superpositions
of states in quantum mechanics \cite{macroQM}, and (ii) the clash
between Quantum Mechanics and General Relativity \cite{GR-QM}, which
has yet to be resolved - the success and generality of each of these
theories makes a resolution of this clash perhaps the most important
single problem in physics today.

\begin{figure}
\vspace{20mm}
\includegraphics[width=8.5cm]{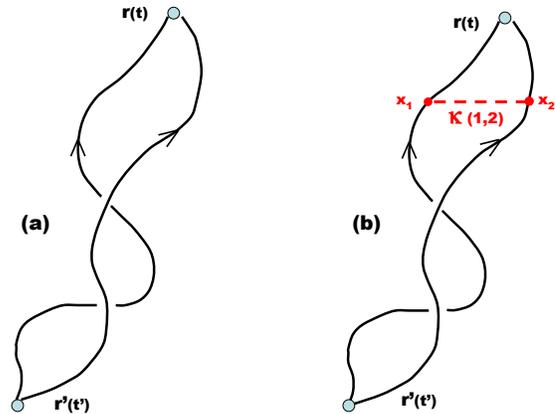}
\vspace{-24mm} \caption{(Color on-line)  Contributions to ${\cal
G}(R, R')$ from different paths. In (a) we show the contributions in
conventional quantum mechanics - paths sum independently. In (b) we
show the intrinsic decoherence contribution, in which the correlator
$\kappa(1,2)$ connects different branches of the propagator}.
 \label{2path}
 \vspace{-4mm}
\end{figure}

As noted above, most previous discussions of the possible breakdown
of quantum mechanics have started from the Schrodinger equation, and
added extra non-linear terms. In what follows I will very briefly
sketch a different framework. The purpose of this sketch is not to
give anything like a full exposition of this theory, for which there
is no space here, but simply to indicate how this theoretical
framework can be set up. Full details will appear elsewhere
\cite{stamp12a}.

Our choice of theoretical framework is motivated by (i) the
observation that the quantum phase, via its connection to the
action, plays a fundamental role in the time evolution of all known
physical systems; and (ii) that the transition amplitude (or just
the 'quantum amplitude') plays a similarly fundamental role. It is
hard to see how these would no longer be incorporated in any new
theory intended to go beyond quantum mechanics. The emphasis on the
quantum phase also focusses attention on phase coherence.

Our starting point will be the standard Feynman path integral
formulation of quantum mechanics, and in this short paper we will
discuss only the non-relativistic version of this, for ordinary
particles. The great advantage of the path integral formulation is
that it makes explicit the idea that at any particular point on its
worldline, a particle can choose between all possible paths. We will
now generalize this idea as follows. We define a propagator ${\cal
G}(R,R')$ for a quantum system between two spacetime points $R' =
({\bf r}', t')$ and $R = ({\bf r},t)$, which we will assume to have
the form
\begin{equation}
{\cal G}(R, R') \;=\; G_o(R, R') + \Delta {\cal G}(R, R')
 \label{Gmod}
\end{equation}
where $G_o({\bf r}, {\bf r}'; t, t')$ is given by the usual path
integral expression in non-relativistic quantum theory, viz.,
\begin{equation}
G_o({\bf r}, {\bf r}'; t, t') \;=\; \int^{{\bf r}}_{{\bf r}'} {\cal
D} {\bf x}(\tau) \exp {i \over \hbar} \int^{t}_{t'} d\tau L({\bf x},
\dot{{\bf x}}; \tau)
 \label{Gtrad}
\end{equation}
whereas the new term $\Delta {\cal G}({\bf r}, {\bf r}'; t, t')$
will be taken, in the first instance, to have the form:
\begin{widetext}
\begin{equation}
\Delta {\cal G}({\bf r}, {\bf r'}; t, t') \;=\;\int^{{\bf r}}_{{\bf
r'}} {\cal D}{\bf x}_1(\tau) \int^{{\bf r}}_{{\bf r}'} {\cal D}{\bf
x}_2(\tau)\;\; \kappa[{\bf x}_1,{\bf x}_2] \exp {i \over
 2\hbar} \int^{t}_{t'} d\tau \; \left[L({\bf x}_1, \dot{{\bf x}}_1;\tau)
 \;+\; L({\bf x}_2, \dot{{\bf x}}_2; \tau) \right]
 \label{G-K}
\end{equation}
\end{widetext}
in which the interpath correlation functional $\kappa[{\bf x}_1,{\bf
x}_2]$ (a functional of the two paths ${\bf x}_1(t), {\bf x}_2 (t)$)
remains to be determined. The above definition is heuristic in the
same way that the usual path integral is, in that the normalization
has yet to be determined (ie., the measure for the paths has yet to
be defined): this point is discussed below. The system is clearly
non-relativistic - we have separated space and time, and the
correlator $\kappa[1,2]$ depends only on integrals over the spatial
variables ${\bf x}_1,{\bf x}_2$, so that the correlations it
mediates are instantaneous.

The correction term $\Delta {\cal G}(R,R')$ describes {\it pairs} of
paths, and it allows communication or correlations {\it between
different paths}. The physical interpretation of (\ref{G-K}) is {\it
not} that that of a self-interaction - in conventional quantum
mechanics, self-interactions for a particle are described by single
particle paths, in which an interaction leaves and then rejoins the
particle path - one then sums over the different possible paths.
Here we are summing over all possible {\it pairs} of paths between
the initial and final state, with the correlator $\kappa[1,2]$
weighting pairs of paths (see Fig. \ref{2path}); the total
transition amplitude is now produced by summing not only over paths
but over correlations between them. We can if we wish think of this
as some kind of communication between different branches of the
wave-function. Clearly there is no particular reason for stopping at
pairs of paths - indeed it would be more elegant to assume a sum
over all possible $n$-tuples of paths, with $0 < n < N$ and $N
\rightarrow \infty$. However we will not discuss this more general
formulation any further here, since we are simply going to assume
for the moment that $K[2,1]$ is very small, so that we can to first
approximation ignore all higher $n$-tuples. We then write
\begin{equation}
\kappa [1,2] = \gamma J[1,2]; \;\;\;\;\;\;\;\;\; (\gamma \ll 1)
 \label{k-K}
\end{equation}
where $\gamma$ is dimensionless and we note that both $J[1,2]$ and
$\kappa[1,2]$ have dimensions of inverse length (or inverse interval
in a relativistic version). We will also typically assume that
$J[1,2]$ takes the form
\begin{equation}
J[x_1,x_2] = \exp \;i \chi[x_1, x_2]
 \label{chi-12}
\end{equation}
where the "phase" $\chi[1,2]$ will in general be complex.

In this theoretical framework, we will assume that both the
amplitude $\psi({\bf r},t)$ and the density matrix still appear;
however the normalization of the path integral is not so obvious,
and in fact has to be handled rather carefully. The amplitude obeys
the equation of motion
\begin{equation}
\psi({\bf r},t) = \int d{\bf r'}{\cal G}({\bf r}, {\bf r}'; t, t')
\; \psi({\bf r'},t')
 \label{eom}
\end{equation}
However the density matrix is no longer given by $\rho({\bf r_1},
{\bf r_2}, t) = \psi({\bf r_1},t) \psi^{\dagger}({\bf r_2},t)$. This
is because we must allow correlations between different paths of the
density matrix, in line with the assumptions made above for the
1-particle propagator. This means that we can still write the
equation of motion for the density matrix $\rho(X,Y) = \rho({\bf x},
{\bf y}, t)$ in the form $\rho(2) = \int d1 {\cal K}(2,1) \rho(1)$
as before, but now the propagator ${\cal K}(X,Y;X'Y')$ contains
correlations between all possible paths connecting the end points
$X,Y,X',Y'$. If we then make the same expansion in powers of
$\lambda$ that is implicit in (\ref{G-K}) above, we get an
expression for the propagator of the form
\begin{equation}
{\cal K}(X,Y;X'Y') = \bar{K}(X,Y;X'Y') + \Delta {\cal K}(X,Y;X'Y')
 \label{K-K}
\end{equation}
where $\bar{K}(X,Y;X'Y') = {\cal G}(X,X') {\cal G}^*(Y,Y')$ is the
product of the two independent propagators (each containing internal
corrections from the correlator $\kappa[1,2]$, as above), and the
correction $\Delta {\cal K}(X,Y;X'Y')$ contains cross-correlations
between the forward and reverse paths in the density matrix
propagator, with the lowest-order term having the expression
\begin{widetext}
\begin{equation}
\Delta {\cal K}(X,Y;X'Y') \;=\;\int^{{\bf X}}_{{\bf X'}} {\cal
D}{\bf x}(\tau) \int^{{\bf Y}}_{{\bf Y'}} {\cal D}{\bf y}(\tau)\;\;
\kappa[{\bf x},{\bf y}] \exp {i \over
 \hbar} \int^{t}_{t'} d\tau \; \left[L({\bf x}, \dot{{\bf x}};\tau)
 \;-\; L({\bf y}, \dot{{\bf y}}; \tau) \right]
 \label{K-corr}
\end{equation}
\end{widetext}
and so on. This last term, because it correlates the 2 parts of the
density matrix propagator, can then have the effect of generating
what looks like decoherence.

One can also define a reduced density matrix in the case where one
deals with a system for which, for one reason or another, it is
necessary to average over some of the degrees of freedom. Suppose
the total density matrix (for the 'universe') is $\rho ({\bf Q},
{\bf Q'}, t)$, where the coordinates ${\bf Q} = ({\bf R}, \{ x_k
\})$, with $k= 1,2,...N$, and we are only interested in the 'central
coordinate ${\bf R}$. Then we simply write that the reduced density
matrix is given as
\begin{eqnarray}
\bar{\rho}({\bf R_1}, {\bf R_2}, t) &=& Tr_{[Env]} \; \rho ({\bf Q},
{\bf Q'}, t) \nonumber \\
&=& \prod_k \int dx_k \; \rho({\bf Q}, {\bf Q'}, t)
 \label{TrX}
\end{eqnarray}
in the usual way.

We may then, by analyzing the dynamics of the reduced density
matrix, examine the corrections to its dynamics coming from the
extra term $\Delta {\cal K}(2,1)$ in the propagator of the full
density matrix. Suppose, for example, that we have a Lagrangian of
form ${\cal L} ({\bf R}, \{ x_k \}) = L_o ({\bf R}) + L_f ({\bf R},
\{ x_k \})$, where the 'fast' Lagrangian is written
\begin{equation}
L_f ({\bf R}, \{ x_k \}) =  L_{\phi}(\{ x_k \}) + L_{int}({\bf R},
\{ x_k \})
 \label{L-f}
\end{equation}
with the fast variables now playing the role of the environment.
Then we can also make such a separation in the propagator ${\cal
G}(2,1)$. The 'bare' term has the standard form \cite{BogSh59}
\begin{equation}
 G_o(2,1) = \int^2_1 {\cal D} {\bf R}\; e^{{i \over \hbar} \int dt L_o ({\bf R},t)}
 G_o^f (2,1)
 \label{G-o-f}
\end{equation}
where $G_o^f(2,1) = G_o^f (\{ x_k^{(2)}, x_k^{(1)} \}; t_2, t_1 | [
{\bf R}(t)])$ is the propagator for the fast variables, conditional
on some path ${\bf R}(t)$ for the slow 'central' variable (ie., the
propagator for the $\{ x_k (t) \}$ when the slow variable path is
'frozen' to be some specific ${\bf R}(t)$). Typically one sandwiches
the fast propagator between the instantaneous eigenstates $|n({\bf
R}(t)) \rangle$ of the entire collection of fast variables, and then
we can write that
\begin{equation}
G_o^f(2,1) = \langle 2 | n ({\bf R}) \rangle G_o^f(m,n; {\bf R}(t))
\langle m({\bf R}) | 1 \rangle
 \label{Gf-nm}
\end{equation}
where $G_o^f(n,m; {\bf R}(t)) = e^{ {\i \over \hbar} \int^t dt'
L_{nm}^f ({\bf R}(t')}$, with an effective Lagrangian
\begin{equation}
L_{nm}^f ({\bf R}(t')) = L_o({\bf R}) - \epsilon_n({\bf R}) - i\hbar
{\bf \dot{R}} \cdot \langle n | \nabla_{\bf R} m \rangle
 \label{L-nm}
\end{equation}
containing the usual 'Berry phase' term derived by Born and
Oppenheimer.

Now suppose we add the extra term $\Delta {\cal G}(2,1)$ to the
propagator. If we again make the separation between slow and fast
variables, we get an extra phase term in the dynamics of the Green
function, and similar extra terms in the propagator for the density
matrix, whose effect is to correlate the variables between 2
separate paths, for both the fast and slow variables. Substituting
(\ref{L-f}) into (\ref{G-K}), we find that these phase factors have
the form
\begin{widetext}
\begin{equation}
\Phi_{nm} [{\bf R}, {\bf R'}] \;=\; \int dt \; {\bf \dot{R}} \cdot
\langle n({\bf R}) | \nabla_{\bf R} | \alpha({\bf R}) \rangle \;
\chi_{\alpha \beta}[{\bf R}, {\bf R'}] \; \langle \alpha({\bf R'}) |
\nabla_{\bf R'} | m({\bf R'}) \rangle \cdot {\bf \dot{R}'}
 \label{Phi-nm}
\end{equation}
\end{widetext}
where $\chi_{\alpha \beta}[{\bf R}, {\bf R'}]$ is given by
\begin{equation}
\chi_{\alpha \beta}[{\bf R}, {\bf R'}] \;=\; | \alpha({\bf R})
\rangle \; \kappa [{\bf R}, {\bf R'}] \; \langle \beta({\bf R'}) |
 \label{XXXXX}
\end{equation}
In analyzing a macroscopic system with many degrees of freedom, this
kind of technique is essential.

One may also look at various limiting cases in this formalism. The
semiclassical limit $\hbar \rightarrow 0$ can be studied directly in
(\ref{G-K}), and we see that since $\hbar$ appears in the
denominator of phase factors, the same subtleties arise as for
ordinary quantum mechanics. In the same way one may study the
adiabatic limit, both for a single particle and for an $N$-particle
system, and the 'macroscopic' limit, where the number of particles
in some object is taken to be very large.

Up until now I have said nothing about how we are supposed to
interpret the density matrix, and expressions like $\langle \hat{A}
\rangle = Tr \; \{ \rho A \}$ (in standard quantum mechanics of
course these refer to expectation values and lead, via the Born
rule, to probabilities for these). Quite apart from these
interpretational questions, we see that there must be a
normalization requirement on the functional integrals - this is of
course a problem even in the path integral formulation of ordinary
quantum mechanics. In the present case the normalization (and the
interpretation of the interpath correlation) depend on what specific
form we employ for the kernel $\kappa (2,1)$; this has then to be
determined for each case. This should not be too surprising -
indeed, the measure of paths in the path integral in ordinary
quantum mechanics also depends on which specific Lagrangian we
choose.  To give a full discussion of this would take us into too
long a technical diversion \cite{stamp12a}.

At first the modifications of ordinary quantum mechanics seem rather
small, if the kernel $\kappa (1,2)$ is assumed to be small. However
the effect on the dynamics of any quantum system actually depends
critically on the form of the kernel $\kappa (1,2)$. It is then
interesting to explore one example, which is of considerable
interest in its own right. Thus we now turn to the specific
mechanism of gravitational decoherence, in which a correlation
between paths mediated by gravitational interactions is assumed to
arise

\section{VI. An Example: Gravitational Decoherence}

Our example of an intrinsic decoherence mechanism is taken from
studies in quantum gravity. Models such as the one to be discussed
have to be taken seriously, given the apparent incompatibility of
between Quantum Mechanics and General Relativity. However I
immediately emphasize that what follows is not intended to do more
than indicate, by way of a toy example, the sort of thing that might
be done - it is very far from being any kind of polished theory.

The basic physical idea to be discussed is that gravitational
effects must lead inevitably to some sort of intrinsic decoherence
in the time evolution of isolated systems. We see that any such
arguments lead us inevitably to a non-unitary time evolution for
state vectors - no averaging over an environment is involved, and
yet phase correlations are lost. Such proposals have been quite
common in the literature, ever since Hawking proposed that such
non-unitary time evolution ought to be a part of any quantum theory
of gravity \cite{hawking82}. Some of these proposals have
incorporated quantum spacetime fluctuations at the Planck energy as
the source of this intrinsic decoherence \cite{ellis}, even to the
extent of estimating how these might cause apparent decoherence in
the dynamics of a superconducting SQUID. These theoretical proposals
are controversial \cite{banks,unruh-wald}, and there has certainly
been no clear experimental test of them at all (note that the
estimates given in the theory for intrinsic decoherence rates
contain a number of factors $\sim O(1)$, which then end up being
exponentiated).

Somewhat distinct from these 'Planck scale' proposals are the ideas
discussed by Penrose \cite{penrose}, following on from Diosi
\cite{diosi}. These authors argue that in analyzing quantum
superpositions of different spacetimes, coming from situations where
a mass is in two different positions, a kind of 'gravitational
time-energy uncertainty principle' must operate. In these analyses,
the timescale $\Delta t_G = \hbar/\Delta E_G$ plays the role of a
decoherence time; here $\Delta E_G$ is the energy scale associated
with the gravitational self-interaction of an object between its
different quantum states. However this decoherence is not
environmental - it can be viewed as either coming from an
uncertainty in the background time, or as arising from a mismatch of
the spacetimes generated by the gravitational field of the system in
question, in its different quantum states.

How can we incorporate these ideas into the formalism above? Let us,
for the sake of having a concrete example, consider a correlator
$\kappa(1,2)$ of form
\begin{equation}
\kappa [{\bf r}, {\bf r'}] \;=\; \exp  \int^t d\tau \;  {4 \pi i G
m^2 \over |{\bf r}(\tau) - {\bf r'}(\tau)|} \;-\;1
 \label{K-G}
\end{equation}
so that the strength of the phase communication between different
paths is now proportional to the strength of the 'gravitational
self-interaction' between them. I make no attempt here to justify
this choice for $\kappa[{\bf r}, {\bf r'}]$ except to note that it
is consistent at least with the spirit of the Diosi-Penrose
uncertainty arguments above. We have assumed that the gravitational
self-interaction can be handled in a Newtonian approximation.

Note, however, one key difference with the simple form introduced
above - the correlator $\kappa[{\bf r}, {\bf r'}]$ is no longer
multiplied by a small parameter $\lambda$. Indeed the phase factor
actually blows up at short distances, even when the mass $m$ is
small; and no power series expansion of the exponential is valid in
the short distance limit.

Consider now a really simple problem, in which we have a free
particle of mass $m$ propagating from $X' = ({\bf R'}, t')$ to $X =
({\bf R}, t)$. In the absence of the correction $\Delta {\cal G}$,
we just have the usual free propagator, ie., ${\cal G}(X,X')
\rightarrow G_o (X,X')$. Adding the gravitational term then gives
the correction
\begin{widetext}
\begin{equation}
\Delta {\cal G}(X,X') \;\propto \; \int {\cal D}{\bf x}_1(\tau) \int
{\cal D}{\bf x}_2(\tau)\; \kappa [{\bf x}_1, {\bf x}_2]
  \exp {i \over 2\hbar} \int d\tau {m \over 2}({\bf {\dot{x}}_1}^2 + {\bf {\dot{x}}_2}^2)
 \label{G-Pen}
\end{equation}
\end{widetext}

This correction affects only the relative coordinate ${\bf r} = {\bf
x}_1 - {\bf x}_2$, and thus we see that
\begin{equation}
\Delta {\cal G}(X,X') \propto {\cal A}(0,0; t,t') G_o(X,X')
 \label{A-G}
\end{equation}
where ${\cal A}({\bf r}, t; {\bf r'},t')$ is just the return
amplitude (starting and finishing at the origin) for a particle of
inertial mass $m$ in an attractive 'Coulomb' field, with a 'charge'
of strength $8 \pi Gm^2$. Thus the main effect on the 1-particle
propagator is just a finite renormalization, whose effects are not
obviously important, since we expect them to be absorbed in the
normalization of the path integral. However the effect on the
density matrix propagator is more dramatic - we have a correction
term which we write out in full, ie.,
\begin{widetext}
\begin{equation}
\Delta {\cal K}(X,Y;X'Y') \;\sim \;\int^{{\bf X}}_{{\bf X'}} {\cal
D}{\bf x}(\tau) \int^{{\bf Y}}_{{\bf Y'}} {\cal D}{\bf y}(\tau)\;\;
\left[ \exp \int d\tau {8 i\pi G m^2 \over |{\bf x}(\tau) - {\bf
y}(\tau)|} \; - 1 \right]
   \times \exp {i \over 2\hbar} \int d\tau {m
\over 2}({\bf {\dot{x}}}^2 - {\bf {\dot{y}}}^2)
 \label{K-corr}
\end{equation}
\end{widetext}
in which decoherence is now caused by what looks like a
gravitational interaction between forward and backward paths of the
density matrix. Indeed, the kernel $\kappa[{\bf x}, {\bf y}]$ is now
acting as an influence functional. Decoherence in this calculation
will appear in the reduced density matrix in both the momentum and
position representations.

An obvious question one can ask about this calculation is that, as
noted above, the intrinsic decoherence correction to the dynamics
here is not small in the limit $|{\bf x}(\tau) - {\bf y}(\tau)|
\rightarrow 0$. Thus we have no reason to believe that these
calculations, limited as they are to the lowest order in the
expansion in $\kappa [1,2]$, give a proper picture of what is
happening. This might be alright if the correlator as given above is
assumed to be exact, but this is hardly likely. Another question,
that has to be dealt with if one wishes to discuss possible
experiments, is to generalize this calculation to deal with a system
of many particles. Thus one can imagine doing an experiment with a
large mass made from particles distributed according to some density
distribution. The way to deal with this is precisely using the
separation of slow and fast variables discussed above - the
calculations are quite lengthy \cite{stamp12a}.

This if course leads us to the key question -- how easy will it be
to see departures from quantum mechanics caused by intrinsic
decoherence? A key problem with many of the theoretical attempts to
predict a breakdown of quantum mechanics is that, because they are
attempting to describe some kind of wave-function collapse, they
give predictions for dynamics which look very much like standard
decoherence. Thus one can ask - how can we distinguish between
intrinsic and environmental decoherence in experiments on, eg., a
macroscopic system like a superconducting SQUID, or an array of
magnetic molecules?

This is clearly a research problem of great interest. On the
experimental side, some very ambitious and interesting efforts exist
to look for gravitational decoherence \cite{bouwm03,aspel12}, in
which the aim is to set up quantum states in which reasonably
massive objects are in superpositions of spatially separated states.
One very nice feature of these experiments is that the gravitational
decoherence rate should indeed vary with external parameters in a
way rather different from what one would reasonably expect from
environmental decoherence rates (at least those coming from
currently understood environmental decoherence mechanisms). However
the problem, at least for tests of gravitational decoherence, is
that there seems to be no unambiguous way of calculating the
gravitational decoherence rate using the simple time-energy
uncertainty arguments that the Diosi-Penrose arguments use. The
answers depend in a quite arbitrary and very sensitive way on how
one 'coarse-grains' the density distribution of the objects involved
- this is simply because of the divergence of the gravitational
'self-energy' term as two mass distributions approach each other, so
that the behaviour of the density distribution at very small
distances determines the intrinsic decoherence rate. One interest of
the above approach is that it may help us to clarify this problem.

It is clearly going to be a very important task for theory to come
up with a clear quantitative prediction for gravitational
decoherence rates - and in fact since experiments of this kind are
perhaps not too far from completion, there may not be too much time
left to do this!

\section{Acknowledgements}

This work was supported by  NSERC, CIFAR, and PITP. I would like to
thank Bill Unruh, Roger Penrose, Dirk Bouwmeester, and Igor Tupitsyn
for discussions of the material.

\vspace{3mm}

$\star$ email address: {\bf stamp@phas.ubc.ca}

\end{document}